\documentstyle[preprint,aps]{revtex}
\tighten
\draft
\widetext
\input epsf
\preprint{HUTP-98/A019, NUB 3175}
\bigskip
\bigskip
\begin{document}
\title{A Three-Family $SU(6)$ Type I Compactification}
\medskip
\author{Zurab Kakushadze\footnote{E-mail: 
zurab@string.harvard.edu}}
\bigskip
\address{Lyman Laboratory of Physics, Harvard University, Cambridge, 
MA 02138\\
and\\
Department of Physics, Northeastern University, Boston, MA 02115}
\date{April 16, 1998}
\bigskip
\medskip
\maketitle

\begin{abstract}
{}We construct a four dimensional chiral ${\cal N}=1$ space-time supersymmetric
Type I vacuum corresponding to a compactification on a toroidal ${\bf Z}_2\otimes {\bf Z}_2
\otimes {\bf Z}_3$ orbifold. Using recent results in four dimensional orientifolds, we argue that
this model has a well defined world-sheet description. An interesting feature of this model
is that the gauge group contains an $SU(6)$ subgroup with three chiral generations. Moreover,
this model contains D5-branes and therefore corresponds to a non-perturbative heterotic vacuum. This is the first example of a consistent 
chiral ${\cal N}=1$ supersymmetric string vacuum which is non-perturbative from the
heterotic viewpoint, has a perturbative description in a dual theory, and
possesses some phenomenologically interesting characteristics. We also compute the tree-level
superpotential in this theory.  
\end{abstract}

{}One of the outstanding questions of string theory is whether it describes {\em our} universe.
It is rather difficult to answer this question as the space of classical string vacua has a very large
degeneracy, and perturbatively there lack objective criteria that would single out a particular string vacuum among the numerous possibilities. It seems reasonable to expect that non-perturbative string dynamics might lift, at least partially, this degeneracy of the moduli space. Understanding non-perturbative string dynamics is therefore of utmost importance. Moreover, in superstring theory supersymmetry is believed to be unbroken perturbatively. 
Since the world around us is not explicitly supersymmetric, understanding non-perturbative mechanisms of supersymmetry breaking in string theory is mandatory if string theory is to
make contact with the real world. 

{}In recent years substantial progress has been made in understanding non-perturbative
string dynamics. In particular, in ten dimensions there are five consistent superstring theories.
The first four, Type IIA, Type IIB, $E_8\otimes E_8$ heterotic and ${\mbox{Spin}}(32)/{\bf Z}_2$
heterotic, are theories of oriented closed strings. Type I superstring is a theory of both unoriented
closed plus open strings. Perturbatively, these five theories are distinct. Nonetheless, these
theories exhibit a web of (conjectured) dualities which all seem to point to an underlying unified
description. Most of these dualities are intrinsically non-perturbative and often allow to
map non-perturbative phenomena in one theory to perturbative phenomena in another theory.

{}Given the success of string dualities in shedding light on non-perturbative string dynamics, it is
natural to wonder whether one can find new consistent string vacua, which are inaccessible perturbatively, such that they possess at least some phenomenologically desirable qualitative
features. Phenomenologically oriented model building in string theory for many years has been 
mostly confined to perturbative heterotic superstring framework where it was facilitated by
the existence of relatively simple rules. In fact, many qualitative features such as realistic gauge groups, number of generations, {\em etc.}, have been successfully obtained within this limited
(as it is now believed to be) 
framework. It is natural to attempt construction of semi-realistic superstring
vacua which would be non-perturbative from the heterotic viewpoint. Not only would this 
be encouraging as far as the hope for making contact with the real world, but it might also
provide additional insight into the structure of string theory and, perhaps, give us clues how
non-perturbative string theory can possibly solve some of the problems which might even seem
hopeless from the perturbative heterotic viewpoint.

{}In the light of the above discussion, one of the promising directions appears to be understanding four dimensional Type I compactifications. One of the reasons to believe that
this might indeed be a fruitful direction stems from the conjectured Type I-heterotic duality \cite{PW}. Thus, the tree-level relation between Type I 
and heterotic dilatons in $D$ space-time dimensions \cite{Sagnotti} 
(which follows from Type I-heterotic duality in ten dimensions \cite{PW}) reads:
\begin{equation}
 \phi_H={{6-D}\over 4} \phi_I -{{D-2}\over 16}\log(\det(g_I))~.
\end{equation} 
Here $g_I$ is the internal metric of the Type I compactification space, 
whereas $\phi_I$ and $\phi_H$ are the Type I and heterotic dilatons, 
respectively. From this one can see that (in four dimensions) 
there always exists a region in the moduli space where both Type I and 
heterotic string theories are weekly coupled, 
so one can hope to be able to understand the duality matching rather precisely.
On the other hand, certain non-perturbative effects on the heterotic side (such as 
dynamics of NS 5-branes which is difficult to study from the heterotic viewpoint) are mapped
to perturbative Type I effects ({\em e.g.}, D5-branes are the Type I duals of heterotic NS 5-branes) which are under much better control.

{}The above expectations seem to be reasonable at the first sight, but there has been
one circumstance preventing substantial progress in these directions. Thus, the four
oriented closed string theories are relatively well understood as far as perturbative formulation is concerned. Conformal field theory and modular invariance serve as guiding principles for 
perturbative model building in closed string theories. Type I, however, lacks modular invariance.
Moreover, conformal field theories on world-sheets with boundaries (invariably present
in open string theories) are still poorly understood. These have been some of the main reasons
for lack of as deep understanding of perturbative Type I compactifications as in closed string theories.

{}In the past years various unoriented closed plus open string vacua have been
constructed using orientifold techniques. Type IIB orientifolds are 
generalized orbifolds 
that involve world-sheet parity reversal along with geometric symmetries 
of the theory \cite{group}. Orientifolding procedure results in an unoriented closed string theory.
Consistency then generically requires introducing open strings that can 
be viewed as starting and ending on D-branes \cite{Db}. 
In particular, Type I compactifications on toroidal orbifolds can be viewed as Type IIB
orientifolds with a certain choice of the orientifold projection. 
Global Chan-Paton charges
associated with D-branes manifest themselves as a gauge symmetry in 
space-time. The orientifold techniques have been successfully
applied to the construction of six dimensional ${\cal N}=1$ space-time supersymmetric 
orientifolds of Type IIB compactified on orbifold limits of K3 (that is, toroidal orbifolds
$T^4/{\bf Z}_N$, $N=2,3,4,6$) \cite{PS,GP,GJ,DP,KST1}. 
These orientifold models generically contain more than one tensor multiplet and/or enhanced gauge symmetries from D5-branes in their massless spectra,
and, therefore, describe six dimensional vacua which are non-perturbative from the 
heterotic viewpoint.

{}The orientifold construction has subsequently been generalized to four dimensional ${\cal N}=1$ space-time supersymmetric compactifications \cite{BL,Sagnotti,ZK,KS1,KS2}. Several
such orientifolds have been constructed. Some of these models, namely, the ${\bf Z}_3$ \cite{Sagnotti}, ${\bf Z}_7$ \cite{KS1} and ${\bf Z}_3\otimes {\bf Z}_3$ \cite{KS2} orbifold
models have perturbative heterotic duals \cite{ZK,KS1,KS2}. Others, such as the ${\bf Z}_2\otimes {\bf Z}_2$ \cite{BL} and ${\bf Z}_6$ \cite{KS2} orbifold models are non-perturbative
from the heterotic viewpoint \cite{ZK,KS1,KS2} as they contain D5-branes. In particular, the  
${\bf Z}_6$ orbifold model of \cite{KS2} is the first known example of a consistent chiral ${\cal N}=1$ string vacuum in four dimensions that is non-perturbative from the heterotic viewpoint.

{}Despite the above developments at some point it became clear that our understanding of 
orientifolds was incomplete. In particular, in some of the models discussed in \cite{Zw,Iba}
the tadpole cancellation conditions (derived using the perturbative orientifold
approach, namely, via a straightforward generalization of the six dimensional 
tadpole cancellation conditions of Refs \cite{PS,GP,GJ,DP} to four dimensions) allowed for no solutions.
This, at least at the first sight, seems surprising as Type IIB compactifications on those
orbifolds are well defined, and so should be the corresponding orientifolds. Clearly, a
better understanding of the orientifold construction was called for.

{}Recently, some progress has been made in this direction \cite{KST}. In particular,
in \cite{KST} conditions necessary for world-sheet
consistency of six and four dimensional ${\cal N}=1$ supersymmetric Type IIB 
orientifolds were studied. It was argued that in most cases orientifolds contain sectors
which are non-perturbative ({\em i.e.}, these sectors have no world-sheet description). These
sectors can be thought of as arising from D-branes wrapping various collapsed 2-cycles
in the orbifold. In particular, such non-perturbative states are present
in the ``anomalous'' models of Ref \cite{Zw} (as well as in other examples of this type
recently discussed in Ref \cite{Iba}). This resolves the corresponding ``puzzles''.
Certain world-sheet consistency conditions in four dimensional cases
(which are automatically satisfied in the six dimensions)
were pointed out in \cite{KST}
which indicate that the only four dimensional orientifolds that have perturbative description
are those of Type IIB compactified on the ${\bf Z}_2\otimes {\bf Z}_2$ \cite{BL}, ${\bf Z}_3$
\cite{Sagnotti}, ${\bf Z}_7$ \cite{KS1}, ${\bf Z}_3\otimes {\bf Z}_3$ and ${\bf Z}_6$
\cite{KS2}, and, finally, ${\bf Z}_2\otimes{\bf Z}_2 \otimes {\bf Z}_3$ (which we study in this paper) orbifolds. In particular,
none of the other models considered in Refs \cite{Zw,Iba} have perturbative orientifold
description, and even in the models with all tadpoles cancelled the massless spectra given
in Refs \cite{Zw,Iba} miss certain non-perturbative states.

{}Although the number of perturbative Type IIB orientifolds in four dimensions appears to be
rather constrained, it is important to understand all the cases at hand. The cases of most interest
are those with D5-branes as they correspond to non-perturbative heterotic vacua. The only orientifold with perturbative description that has not been previously studied is that of Type IIB
on the ${\bf Z}_2\otimes{\bf Z}_2 \otimes {\bf Z}_3$ orbifold. This is the case we will focus on in 
this paper. Happily, we will find a pleasant ``surprise'', namely, the gauge group and matter
content of this model (which is chiral) belong to the class of ``semi-realistic'' models which we were hoping to obtain to begin with. Moreover, this model is non-perturbative from the
heterotic viewpoint (as it contains D5-branes), so it sheds some additional light on non-perturbative dynamics of heterotic NS 5-branes. 

{}The phenomenologically attractive features of this model are the gauge group and the number of chiral families. Thus, the gauge group of this model (for a particular configuration of D-branes)
is $[SU(6)\otimes Sp(4)]^4$. The four copies of the $SU(6)\otimes Sp(4)$ gauge symmetry come from D9- and three sets of D5-branes. If we regard one of the $SU(6)$ subgroups as the observable gauge group that contains the Standard Model, then the number of generations is
three: there are 3 chiral ${\bf 15}$'s of $SU(6)$ plus some number of ${\bf 6}$'s and ${\overline 
{\bf 6}}$'s. (Note that the number of generations in $SU(6)$ is determined by the number of chiral 
${\bf 15}$'s.) Here we point out that in this model the charged matter is such that one cannot 
break the ``grand unified'' $SU(6)$ gauge group down
to the $SU(3)_c\otimes SU(2)_w\otimes U(1)_Y$ gauge group of the Standard Model
by Higgsing. Nonetheless, it is very pleasing to observe that relatively simple Type I compactifications can yield certain qualitative features, such as the gauge group and the number of chiral generations, which are desirable phenomenologically. This is especially satisfying taking into account that
this model is non-perturbative from the heterotic viewpoint. In fact, this is the first example of
a chiral ${\cal N}=1$ supersymmetric string vacuum which is non-perturbative from the
heterotic viewpoint, has a perturbative description in a dual (that is, Type I) theory, and
possesses some nice phenomenological characteristics.

{}Clearly, better understanding of four dimensional compactifications of Type I on more
generic Calabi-Yau spaces appears to be more than desirable. Construction of the model
of this paper brings some hope that we might be able to learn more about phenomenological
implications of non-perturbative heterotic vacua in the light of our discussion in the beginning of this section. Moreover, deepening or understanding of Type I compactifications would certainly
be useful in understanding the underlying unifying theory and the nature of string dualities.
 
\begin{center}
{\bf  Type I on the} ${\bf Z}_2\otimes{\bf Z}_2 \otimes{\bf Z}_3$ {\bf  Orbifold}\\
\end{center}
     
{}In the following we describe the construction of this Type I vacuum in detail.
We compactify ten dimensional
Type I superstring theory on a toroidal orbifold ${\cal M}=T^6/G$ where $G$ is the orbifold
group generated by three elements $g,r_1,r_2$. To describe the action of the orbifold group,
let us for simplicity take $T^6$ to be a product of three two-tori: $T^6=T^2\otimes T^2\otimes
T^2$. Let $z_i$ ($i=1,2,3$) be the complex coordinates parametrizing the $T^2$'s. Then the action of the orbifold group $G$ is given by:
\begin{equation}\label{gen}
 gz_i=\omega z_i~,~~~r_i z_j=-(-1)^{\delta_{ij}} z_j~.
\end{equation}
Here $r_3=r_1 r_2$, and $\omega=\exp(2\pi i/3)$. Note that $r_1$ generates a ${\bf Z}_2$
group. Similarly, $r_2$ generates a different ${\bf Z}_2$ group. Finally, $g$ generates a 
${\bf Z}_3$ group. Moreover, from (\ref{gen}) it is clear that $g,r_1,r_2$ commute, so that
together they generate the orbifold group $G=\{g^k,r_i g^k\vert i,k=1,2,3\}\approx{\bf Z}_3
\otimes{\bf Z}_2\otimes{\bf Z}_2$. (The element $g^3=1$ is the identity element of the group
$G$.)

{}It is convenient to view Type I compactification on ${\cal M}$ as an orientifold of Type IIB
compactified on ${\cal M}$. The orientifold projection is then simply given by $\Omega$, 
where $\Omega$ is the world-sheet parity reversal that interchanges the left- and right-movers
of Type IIB. Before we discuss this orientifold of Type IIB on ${\cal M}$, let us first understand
Type IIB compactification on the ${\cal M}$ orbifold. Note that ${\cal M}$ is a Calabi-Yau three-fold with $SU(3)$ holonomy. This implies that Type IIB on ${\cal M}$ is a four dimensional
theory with ${\cal N}=2$ space-time supersymmetry. Let us compute the 
massless spectrum of this theory. For this it suffices to know the Hodge numbers of ${\cal M}$
Calabi-Yau three-fold. Throughout this paper we will assume that the NS-NS antisymmetric tensor background
is trivial ({\em i.e.}, $B_{ij}=B_{{\bar i}{\bar j}}=B_{i{\bar j}}=0$).
Then it is not difficult to deduce the Hodge numbers 
in the orbifold sense. 
Here we give the contributions from each sector (untwisted and twisted)
into the Hodge numbers $(h^{1,1},h^{2,1})$:
untwisted: $(3,0)$; $g$ plus $g^{-1}$ twisted (that is, ${\bf Z}_3$ twisted): $(9,0)$;
$r_i$ twisted (that is, (${\bf Z}_2)_i$ twisted): $(6,0)$; $gr_i$ plus $g^{-1} r_i$ twisted 
(that is, (${\bf Z}_6)_i$ twisted): $(2,0)$; total $(h^{1,1},h^{2,1})=(36,0)$. 

{}Note that $(h^{1,1},h^{2,1})=(36,0)$ are the Hodge numbers of the $Z$-orbifold, that is,
of the quotient ${\cal M}^\prime=T^6/G^\prime$, where $G^\prime=\{1,g,g^2\}\approx{\bf Z}_3$,
and $g$ acts as in (\ref{gen}) on the complex coordinates $z_i$ on $T^6$. (Here and above we
assume that the corresponding tori possess the symmetry by which we mod out or else
the orbifold construction would be inconsistent.) Thus, ${\cal M}$ and ${\cal M}^\prime$ topologically are the same. However, as we will see below, the perturbative orientifolds
of Type IIB on these two quotients give rise to two different Type I models. (This is due to
the different choices of perturbative (from the orientifold viewpoint) 
gauge bundles in the two cases.) 

{}Next, we turn to the $\Omega$ orientifold of Type IIB on ${\cal M}$. Here some caution is required. The point is that the orientifold projection $\Omega$ must be chosen to be the same 
as in the case of Type IIB on a smooth Calabi-Yau three-fold. Such an orientifold projection
leads to Type I compactification on ${\cal M}$. The reasons why this choice of the orientifold
projection is forced have been recently discussed at length in \cite{KST}. In particular, we do not
have an option of choosing the orientifold projection analogous to that in the six dimensional models of \cite{GJ,DP}. On the other hand, the above $\Omega$ orientifold projection is
not a symmetry of Type IIB on ${\cal M}$ at the orbifold conformal field theory point \cite{KST}.
The reason for this is that $\Omega$ correctly reverses the world-sheet orientation of world-sheet bosonic and fermionic oscillators and left- and right-moving momenta, but fails
to do the same with the {\em twisted} ground states. (Such a reversal would involve mapping  
the $g$ and $gr_i$ twisted ground states to the $g^{-1}$ and $g^{-1}r_i$ twisted ground states, respectively. In \cite{KST} such an orientation reversal was shown to be inconsistent.) This difficulty is circumvented by noting that the orientifold projection $\Omega$ is consistent for
{\em smooth} Calabi-Yau three-folds, and, in particular, for a blown-up version of the ${\cal M}$
orbifold. Thus, once the appropriate blow-ups are performed, the orientifold procedure is well
defined. (In particular, the ${\bf Z}_3$ and $({\bf Z}_6)_i$ twisted sector singularities must be
blown up. The $({\bf Z}_2)_i$ singularities are harmless, however \cite{KST}.)

{}Another important issue is related to the question of whether 
all the sectors of the orientifold have a well defined world-sheet
interpretation. Note that the orientifold group is given by ${\cal O}=
\{g^k,r_i g^k, \Omega g^k, \Omega r_i g^k \vert i,k=1,2,3\}$. The sectors labeled by
$g^k,r_i g^k$ correspond to the (unoriented) closed string sectors. The presence
of the element $\Omega\in{\cal O}$ implies that we must introduce 32 D9-branes (just
at in ten dimensional Type I). Similarly, the presence of the elements $\Omega r_i\in{\cal O}$
requires introduction of three sets of D5-branes, to which we refer to as D$5_i$-branes, with
32 D5-branes in each set. (The fact that we need 32 D5-branes in each set can be seen from 
T-duality between these D5-branes and D9-branes.) All of these sectors have a well defined
world-sheet description in terms of open strings starting and ending on D-branes ({\em i.e.}, they
are perturbative from the orientifold viewpoint). On the other hand, as pointed out in 
\cite{KST}, the sectors corresponding
to the elements $\Omega g^k$ and $\Omega r_i g^k$ ($k=1,2$) do not have interpretation in
terms of open strings stretched between D-branes. In particular, such sectors would correspond
to open strings with mixed (that is, neither Dirichlet nor Neumann) boundary conditions. The corresponding states cannot be described perturbatively. They can be viewed as arising from
D-branes wrapping collapsed two-cycles in the orbifold \cite{KST}. Such states do not have a world-sheet description, that is, they are non-perturbative from the orientifold viewpoint.

{}This difficulty is a generic feature in most of the orientifolds of Type IIB compactified on toroidal
orbifolds. However, there is a (rather limited) class of cases where the would-be non-perturbative
states are massive (and decouple in the low energy effective field theory) if we consider
compactifications on blown-up orbifolds \cite{KST}. The orbifold we are considering here
is precisely one of these examples.
In fact, we have already pointed out that blowing up
the ${\bf Z}_3$ and $({\bf Z}_6)_i$ orbifold singularities is forced on us by the world-sheet
consistency conditions. Note that the structure of the fixed points in the $({\bf Z}_6)_i$ 
twisted sectors
is the same as of some of the fixed points in the ${\bf Z}_3$ twisted sectors. This implies
that blowing up the ${\bf Z}_3$ singularities automatically results in the corresponding 
blow-ups in the $({\bf Z}_6)_i$ twisted sectors. In \cite{ZK} using Type I-heterotic
duality it was shown that all the non-perturbative (from the orientifold viewpoint) states
decouple in the $\Omega$ orientifold of Type IIB on ${\cal M}^\prime$ upon blowing up
${\cal M}^\prime$ to a smooth Calabi-Yau three-fold. This fact was utilized in
\cite{KST} to argue that all the non-perturbative states in the $\Omega$ orientifold of Type IIB on ${\cal M}$ also decouple once we appropriately blow up the ${\bf Z}_3$ singularities in
${\cal M}$. This decoupling can be explicitly seen \cite{ZK} in the heterotic dual 
(which is perturbative from the heterotic viewpoint as there are no D5-branes in the corresponding Type I model) of the $\Omega$ orientifold of Type IIB on ${\cal M}^\prime$.
In particular, on the heterotic side the states corresponding to the $\Omega g^k$ and $\Omega r_i g^k$ ($k=1,2$) elements of the orientifold group ${\cal O}$ are charged matter fields 
arising in the twisted sectors of the heterotic orbifold. These states, which we symbolically
denote by $T_a$, are perturbative from the heterotic viewpoint. (The heterotic dual is a modular
invariant orbifold conformal field theory.) The blow-up modes $S_a$ of the orbifold also come
from the twisted sector states in the heterotic model. Moreover, there is a perturbative superpotential of the form ${\cal W}={\cal Y}_{abc} S_a T_a T_b+\dots$ with the Yukawa
couplings precisely such that the states $T_a$ decouple once the blow-up modes $S_a$ acquire appropriate vevs (so that the orbifold is blown up to a smooth Calabi-Yau three-fold).
In \cite{KST} it was argued that an analogous effect takes place in the $\Omega$ orientifold of Type IIB on ${\cal M}$. All the naively expected non-perturbative (from the orientifold viewpoint)
states therefore decouple once we consider the corresponding orientifold on a blown-up
${\cal M}$.
 
{}The above analyses imply that we can use the ``naive'' tadpole cancellation conditions
(derived by straightforwardly generalizing those in six dimensional cases of Refs \cite{PS,GP,GJ,DP} to four dimensions) to construct the massless spectrum of the
$\Omega$ orientifold of Type IIB on the (blown up) ${\cal M}$ orbifold. The
tadpole cancellation conditions for examples of this type were discussed in \cite{KS1,KS2}
as a generalization of the cases studied in \cite{PS,GP,GJ,DP,BL,Sagnotti}. (Also see  
\cite{Zw,O'D,Iba}.) Here for the sake of brevity we will skip the details and simply state the
answer.

{}The action of the orbifold group $G\equiv\{g_a\vert a=1,\dots,|G|\}$ on the Chan-Paton charges carried by the D9- and D$5_i$-branes is described by $32\times 32$ matrices
$\gamma_{g_a,9}$ and $\gamma_{g_a,5_i}$, respectively. These matrices must form a projective representation of $G$. ({\em A priori} they can also 
form a representation of the double cover of $G$.) Note that the Chan-Paton matrices
$\gamma_{1,9}$ and $\gamma_{1,5_i}$ corresponding to the identity element of $G$ can be
chosen to be $32\times 32$ identity matrices, and ${\mbox{Tr}}(\gamma_{1,9})=n_9$
and ${\mbox{Tr}}(\gamma_{1,5_i})=n_{5_i}$, where $n_9=32$ and $n_{5_i}=32$ are the 
numbers of D9- and D$5_i$-branes, respectively. (The constraints on the numbers of D-branes
arise from the untwisted tadpole cancellation conditions.) The twisted tadpole cancellation conditions read:
\begin{eqnarray}
 &&{\mbox{Tr}}(\gamma_{g,9})={\mbox{Tr}}(\gamma_{g^2,9})=-4~,\\
 &&{\mbox{Tr}}(\gamma_{r_i,9})={\mbox{Tr}}(\gamma_{g r_i,9})=
 {\mbox{Tr}}(\gamma_{g^2 r_i,9})=0~,
\end{eqnarray}
and similarly for the Chan-Paton matrices $\gamma_{g_a,5_i}$. A choice\footnote{This choice is unique up to equivalent representations. The uniqueness of the choice
for ${\mbox{Tr}}(\gamma_{g,9})$ is not difficult to see. On the other hand, the uniqueness
of the choice for ${\mbox{Tr}}(\gamma_{r_i,9})$ was argued in \cite{BL} from the orientifold viewpoint, and was recently shown in \cite{KST} from F-theory \cite{vafa} considerations.}  consistent with the requirement that the Chan-Paton matrices form a (projective) representation (of the double cover) of $G$ is given by:
\begin{eqnarray}
 &&\gamma_{g,9}={\mbox{diag}}(\omega I_{12},\omega^2 I_{12}, I_{8})~,\\
 &&\gamma_{r_i,9}=i\sigma_i\otimes I_{16}~, 
\end{eqnarray}
where $\sigma_i$ are $2\times 2$ Pauli matrices, and $I_M$ is an $M\times M$ identity matrix.
(The action on the D$5_i$ Chan-Paton charges is similar.)

{}It is not difficult to compute the massless spectrum of this theory. First, consider the closed 
string sectors. Before the orientifold projection we have ${\cal N}=2$ four dimensional supergravity coupled to $h^{1,1}=36$ neutral hypermultiplets and $h^{2,1}=0$ vector multiplets.
After the orientifold projection we have ${\cal N}=2$ supergravity coupled to 
$h^{1,1}+h^{2,1}=36$ \cite{KST} neutral chiral multiplets. These states are summarized in Table
\ref{model} sector-by-sector. Next, consider the open string sectors. We have 99, $5_i 5_i$,
$95_i$ and $5_i 5_j$ ($i\not=j$)
open string sectors. We will confine our attention to the vacuum where
the D$5_i$-branes are sitting on top of each other at the origin in ${\cal M}$, and take all the 99
sector Wilson lines to be trivial. Then 99 and $5_i 5_i$ gauge groups are identical.
Thus, the 99 gauge group is $SU(6)\otimes Sp(4)\otimes U(1)$. (Here we are using the convention where $Sp(2M)$ has rank $M$.) The massless fields (chiral multiplets) charged
under the gauge group are summarized in Table \ref{model}. Note that in each $SU(6)$ subgroup
we have 3 chiral generations. (The number of generations in $SU(6)$ is determined by the
number of chiral representations transforming in ${\bf 15}$ of $SU(6)$.)

{}Here we should point out that the 99 and $5_i 5_i$ $U(1)$'s are all anomalous \cite{anom}
in this model. We therefore have the corresponding Fayet-Iliopoulos D-terms. These D-terms 
are cancelled via a generalized Green-Schwarz mechanism \cite{GS}. There is a subtlety here,
however. The blow-up modes (that is, the neutral chiral multiplets in the closed string sector)
transform non-trivially under the anomalous U(1) gauge transformation. (This was suggested in \cite{Sagnotti} and shown in \cite{ZK} explicitly in the case of the $\Omega$ orientifold
of Type IIB on ${\cal M}^\prime$ using Type-I heterotic duality.) These blow-up modes
(together with the dilaton plus axion supermultiplet) cancel the Fayet-Iliopoulos D-terms
via acquiring vevs. This is consistent with the observation that the orbifold singularities
must be blown up \cite{KST} as we discussed previously. In fact, from Type I-heterotic duality
one can see \cite{ZK} that it is precisely the ${\bf Z}_3$ blow-up modes that are responsible
for cancelling the D-terms in accord with our previous discussions. As a result, all the $U(1)$'s are broken (and must therefore be deleted from the massless spectrum in Table \ref{model}).
Moreover, 4 out of the 36 chiral multiplets in the closed string sector are eaten in the super-Higgs mechanism.
  
{}This Type I vacuum contains D5-branes and therefore its heterotic dual is non-perturbative:
under Type I-heterotic duality Type I D5-branes are mapped to heterotic NS 5-branes. Let us
understand the dynamics of D5-branes in more detail. In particular, we would like to study flat
directions corresponding to motion of D5-branes. Since we are dealing with a four dimensional ${\cal N}=1$ vacuum, we need to deduce the (tree-level) superpotential in this model. 
It can be computed using the $H$-charge conservation rule \cite{Hamidi}. The $H$-charges are given in Table \ref{model}. Here we give the non-vanishing renormalizable terms only (we suppress the actual values of the Yukawa couplings):
\begin{eqnarray}
 {\cal W}=&&\epsilon_{ijk} \Phi_i\chi_j\chi_k+\epsilon_{ijk} \Phi^l_i\chi^l_j\chi^l_k
 +\epsilon_{ijk} \Phi^i_k Q^{ij} Q^{ij} + \epsilon_{ijk} \chi^i_k P^{ij} R^{ij}+\nonumber\\
 && \Phi_i Q^{i} Q^{i} + \chi_i P^{i} R^{i} +P^{ij} Q^{jk} R^{kj}+\nonumber\\
 \label{sup}
 &&Q^{ij} P^i Q^j+R^{ij} Q^i P^j+P^{ij} R^i R^j+\dots~.
\end{eqnarray}
Here summation over repeated indices is understood. The F-flatness conditions can be read off from this superpotential, and together with the D-flatness conditions they determine the classical
moduli space of the theory.

\begin{center}
{\bf  Acknowledgments}\\
\end{center}

{}I would like to thank Pran Nath, Gary Shiu, Henry Tye and Edward Witten for
discussions. This work was supported in part by the grant NSF PHY-96-02074, 
and the DOE 1994 OJI award. I would also like to thank Albert and Ribena Yu for 
financial support.

\begin{table}[t]
\begin{tabular}{|c|l|l|l|l|}
 Sector & Field & $[SU(6)\otimes Sp(4) \otimes U(1)]^4$
        & $(H_l)_{-1}$ & $(H_l)_{-1/2}$ \\
\hline
Closed & & & &\\
Untwisted & & 3   & & \\
\hline
Closed &  & & & \\
${\bf Z}_3$ Twisted & & 9 & & \\
\hline
Closed & & & &  \\
$({\bf Z}_6)_i$ Twisted & & $3\times 2$ & & \\
\hline
Closed & & & &  \\
$({\bf Z}_2)_i$ Twisted & & $3\times 6$ & & \\
\hline
      & $\Phi_k$    & $3\times ({\bf 15},{\bf 1})
(+2)_L$ & $\delta_{kl}$ & $\delta_{kl}-{1\over 2}$ \\
Open $99$      & $\chi_k$    & $3\times({\overline {\bf 6}},{\bf 4})
(-1)_L$ & $\delta_{kl}$ & $\delta_{kl}-{1\over 2}$ \\
\hline
      & $\Phi^i_k$    & $3\times ({\bf 15}_i,{\bf 1}_i)
(+2_i)_L$ & $\delta_{kl}$ & $\delta_{kl}-{1\over 2}$ \\
Open $5_i5_i$      & $\chi^i_k$    & $3\times ({\overline {\bf 6}}_i,{\bf 4}_i)
(-1_i)_L$ & $\delta_{kl}$ & $\delta_{kl}-{1\over 2}$ \\
\hline
      & $P^i$    & $({\bf 6},{\bf 1};{\bf 6}_i,{\bf 1}_i)
(+1;+1_i)_L$ & ${1\over 2}-{1\over 2}\delta_{il}$ & $-{1\over 2}\delta_{il}$ \\
Open $95_i$      & $Q^i$    & $({\overline {\bf 6}},{\bf 1};{\bf 1}_i,{\bf 4}_i)(-1;0_i)_L$ & ${1\over 2}-{1\over 2}\delta_{il}$ & $-{1\over 2}\delta_{il}$ \\
& $R^i$    & $({\bf 1},{\bf 4};{\overline {\bf 6}}_i,{\bf 1}_i)(0;-1_i)_L$ & ${1\over 2}-{1\over 2}\delta_{il}$ & $-{1\over 2}\delta_{il}$ \\
\hline
      & $P^{ij}$    & $({\bf 6}_i,{\bf 1}_i;{\bf 6}_j,{\bf 1}_j)
(+1_i;+1_j)_L$ & ${1\over 2}-{1\over 2}|\epsilon_{ijl}|$ & $-{1\over 2}|\epsilon_{ijl}|$ \\
Open $5_i 5_j$      & $Q^{ij}$    & $({\overline {\bf 6}}_i,{\bf 1}_i;{\bf 1}_j,{\bf 4}_j)(-1_i;0_j)_L$& ${1\over 2}-{1\over 2}|\epsilon_{ijl}|$ & $-{1\over 2}|\epsilon_{ijl}|$ \\
& $R^{ij}$    & $({\bf 1}_i,{\bf 4}_i;{\overline {\bf 6}}_j,{\bf 1}_j)(0_i;-1_j)_L$ & ${1\over 2}-{1\over 2}|\epsilon_{ijl}|$ & $-{1\over 2}|\epsilon_{ijl}|$ \\
\hline
\end{tabular}
\caption{The massless spectrum of the four dimensional 
Type I ${\bf Z}_3\otimes{\bf Z}_2\otimes {\bf Z}_2$
orbifold model with ${\cal N}=1$ space-time 
supersymmetry and gauge group 
$[SU(6)\otimes Sp(4) \otimes U(1)]^4 $. The number of neutral chiral multiplets in
each of the closed string sectors is shown.
The $H$-charges in both $-1$ and $-1/2$ pictures for states
in the open
string sector are also given. The gravity, dilaton and gauge supermultiplets 
are not shown. The indices $i,j,k,l$ take values $1,2,3$.}  
\label{model}
\end{table}


\begin{references}

\bibitem{PW} J. Polchinski and E. Witten, Nucl. Phys. {\bf B460} (1996) 525, hep-th/9510169.

\bibitem{Sagnotti} C. Angelantonj, M. Bianchi, G. Pradisi, A. Sagnotti and 
Ya.S. Stanev, Phys. Lett. {\bf B385} (1996) 96, hep-th/9606169.

\bibitem{group} S.M. Roy and V. Singh, Phys. Rev. {\bf D35} (1987) 1939; 
Phys. Rev. {\bf D36} (1987) 1827;\\
C.G. Callan, C. Lovelace, C.R. Nappi, S.A. Yost,
Nucl. Phys. {\bf B293} (1987) 83; Nucl. Phys. {\bf B308} (1988) 221;\\
A. Sagnotti, in Cargese'87, eds. G. Mack  {\em et al.} (Pergamon Press, 1988) p. 521;\\
J. Polchinski and Y. Cai, Nucl. Phys. {\bf B296} (1988) 91;\\
N. Ishibashi and T. Onogi, Nucl. Phys. {\bf B318} (1989) 239;\\
D.C. Dunbar, Nucl. Phys. {\bf B319} (1989) 72;\\
Z. Bern and D.C. Dunbar, Nucl. Phys. {\bf B319} (1989) 104;\\
P. Ho{\u r}ava, Nucl. Phys. {\bf B327} (1989) 461; Phys. Lett. {\bf B231} 
(1989) 251;\\
J. Dai, R.G. Leigh and J. Polchinski, Mod. Phys. Lett. {\bf A4} (1989) 2073;\\
R.G. Leigh, Mod. Phys. Lett. {\bf A4} (1989) 2767.

\bibitem{Db} J. Polchinski, Phys. Rev. Lett. {\bf 75} (1995) 4724.

\bibitem{PS} G. Pradisi and A. Sagnotti, Phys. Lett. {\bf B216} (1989) 59;\\
M. Bianchi and A. Sagnotti, Phys. Lett. {\bf B247} (1990) 517; Nucl. Phys. 
{\bf B361} (1991) 539. 

\bibitem{GP} E.G. Gimon and J. Polchinski, Phys. Rev. {\bf D54} (1996) 1667, 
hep-th/9601038.

\bibitem{GJ} E.G. Gimon and C.V. Johnson, Nucl. Phys. {\bf B477} (1996) 715, 
hep-th/9604129.

\bibitem{DP} A. Dabholkar and J. Park, Nucl. Phys. {\bf B477} (1996) 701, 
hep-th/9604178.

\bibitem{KST1} Z. Kakushadze, G. Shiu and S.-H.H. Tye, hep-th/9803141.

\bibitem{BL} M. Berkooz and R.G. Leigh, Nucl. Phys. {\bf B483} (1997) 187,
hep-th/9605049.

\bibitem{ZK} Z. Kakushadze, Nucl. Phys. {\bf B512} (1998) 221, hep-th/9704059.

\bibitem{KS1} Z. Kakushadze and G. Shiu, Phys. Rev. {\bf D56} (1997) 3686,
hep-th/9705163.

\bibitem{KS2} Z. Kakushadze and G. Shiu, hep-th/9706051.

\bibitem{Zw} G. Zwart, hep-th/9708040.

\bibitem{Iba} G. Aldazabal, A. Font, L.E. Ib{\'a}{\~n}ez and
G. Violero, hep-th/9804026.

\bibitem{KST} Z. Kakushadze, G. Shiu and S.-H.H. Tye, hep-th/9804092.

\bibitem{O'D} D. O'Driscoll, hep-th/9801114.

\bibitem{vafa} C. Vafa, Nucl. Phys. {\bf B469} (1996) 403, hep-th/9602022.

\bibitem{anom} E. Witten, 
Phys. Lett. {\bf B149} (1984) 351;\\
M. Dine, N. Seiberg and E. Witten,
Nucl. Phys. {\bf B289} (1987) 589.

\bibitem{GS} M. Green and J.H. Schwarz,
Phys. Lett. {\bf B149} (1984) 117; Phys. Lett. {\bf B151} (1985) 21.

\bibitem{Hamidi} S. Hamidi and C. Vafa, Nucl. Phys. {\bf B279} (1987) 465; \\
L. Dixon, D. Friedan, E. Martinec and S. Shenker, Nucl. Phys.
{\bf B282} (1987) 13.

\end{references}
\end{document}